%% file: main.tex
\documentclass[acmsmall]{acmart}

\AtBeginDocument{%
  \providecommand\BibTeX{{%
    \normalfont B\kern-0.5em{\scshape i\kern-0.25em b}\kern-0.8em\TeX}}}

\setcopyright{acmcopyright}
\copyrightyear{2018}
\acmYear{2018}
\acmDOI{10.1145/1122445.1122456}

\acmJournal{JACM}
\acmVolume{37}
\acmNumber{4}
\acmArticle{111}
\acmMonth{8}

\usepackage{wrapfig}
\usepackage{enumitem}
\usepackage{tikz}
\usepackage{todonotes}
\usepackage[flushleft]{threeparttable}

\usepackage{amsfonts,amsmath,amssymb,array} % choose if needed
\usepackage{multirow}

\usepackage{array}
\newcolumntype{L}{>{\centering\arraybackslash}m{2.2cm}}
\newcolumntype{J}{>{\centering\arraybackslash}m{2.6cm}}
\newcolumntype{K}{>{\centering\arraybackslash}m{2.5cm}}
\newcolumntype{C}{>{\centering\arraybackslash}m{1.7cm}}

\usepackage{booktabs}

\newcommand{\arv}{\textcolor{black}}
\newcommand{\ebv}{\textcolor{black}}
\newcommand{\nsv}{\textcolor{black}}
\newcommand{\seimav}{\textcolor{black}}

\begin{document}

\title{A Survey on Time-Sensitive Resource Allocation in the Cloud Continuum}

\author{Saravanan Ramanathan}
\orcid{0000-0002-1894-4195}
%\email{webmaster@marysville-ohio.com}
\affiliation{%
	\institution{TUMCREATE Ltd.}
\streetaddress{1 Create Way}
\country{Singapore}
\postcode{138602}}
\email{saravanan.ramanathan@tum-create.edu.sg}

\author{Nitin Shivaraman}
\affiliation{%
	\institution{TUMCREATE Ltd.}
\streetaddress{1 Create Way}
\country{Singapore}
\postcode{138602}}
\email{nitin.shivaraman@tum-create.edu.sg}

\author{Seima Suryasekaran}
\affiliation{%
	\institution{TUMCREATE Ltd.}
\streetaddress{1 Create Way}
\country{Singapore}
\postcode{138602}}
\email{seima.suriyasekaran@tum-create.edu.sg}

\author{Arvind Easwaran}
\affiliation{%
 \institution{Nanyang Technological University}
 \streetaddress{50 Nanyang Avenue}
 \country{Singapore}
   \postcode{639798}}
\email{arvinde@ntu.edu.sg}

\author{Etienne Borde}
\affiliation{%
  \institution{T{\'e}l{\'e}com Paris}
  \streetaddress{19 Place Marguerite Perey}
  \city{Palaiseau}
  \country{France}
    \postcode{91120}}
\email{etienne.borde@telecom-paris.fr}

\author{Sebastian Steinhorst}
\affiliation{%
  \institution{Technische Universit{\"a}t M{\"u}nchen}
  \streetaddress{90 Theresienstra{\ss}e}
  \city{M{\"u}nchen}
  \country{Germany}
  \postcode{80333}}
\email{sebastian.steinhorst@tum.de}

\renewcommand{\shortauthors}{Ramanathan, et al.}

\begin{abstract}
  Artificial Intelligence (AI) and Internet of Things (IoT) applications are rapidly growing in today's world where they are continuously connected to the internet and process, store and exchange information among the devices and the environment. The cloud and edge platform is very crucial to these applications due to their inherent compute-intensive and resource-constrained nature. One of the foremost challenges in cloud and edge resource allocation is the efficient management of computation and communication resources to meet the performance and latency guarantees of the applications. 
  The heterogeneity of cloud resources (processors, memory, storage, bandwidth), variable cost structure and unpredictable workload patterns make the design of resource allocation techniques complex.
  Numerous research studies have been carried out to address this intricate problem.
  In this paper, the current state-of-the-art resource allocation techniques for the cloud continuum, in particular those that consider time-sensitive applications, are reviewed. 
  Furthermore, we present the key challenges in the resource allocation problem for the cloud continuum, a taxonomy to classify the existing literature and the potential research gaps.
\end{abstract}




\maketitle

\input introduction
\input model
\input{classification_criteria}
\input{survey}
\input conclusion

\begin{acks}
This work was financially supported in part by the Singapore National Research Foundation under its Campus for Research Excellence And Technological Enterprise (CREATE) programme.
\end{acks}

\bibliographystyle{ACM-Reference-Format}
\bibliography{references}

%\appendix
%Lorem ipsum dolor sit amet, consectetur adipiscing elit. Morbi
%malesuada, quam in pulvinar varius, metus nunc fermentum urna, id
%sollicitudin purus odio sit amet enim. Aliquam ullamcorper eu ipsum
%vel mollis. Curabitur quis dictum nisl. Phasellus vel semper risus, et
%lacinia dolor. Integer ultricies commodo sem nec semper.
%Etiam commodo feugiat nisl pulvinar pellentesque. Etiam auctor sodales
%ligula, non varius nibh pulvinar semper. Suspendisse nec lectus non
%ipsum convallis congue hendrerit vitae sapien. Donec at laoreet
%eros. Vivamus non purus placerat, scelerisque diam eu, cursus
%ante. Etiam aliquam tortor auctor efficitur mattis.
%Nam id fermentum dui. Suspendisse sagittis tortor a nulla mollis, in
%pulvinar ex pretium. Sed interdum orci quis metus euismod, et sagittis
%enim maximus. Vestibulum gravida massa ut felis suscipit
%congue. Quisque mattis elit a risus ultrices commodo venenatis eget
%dui. Etiam sagittis eleifend elementum.
%
%Nam interdum magna at lectus dignissim, ac dignissim lorem
%rhoncus. Maecenas eu arcu ac neque placerat aliquam. Nunc pulvinar
%massa et mattis lacinia.

\end{document}

%% file: introduction.tex
\section{Introduction}
\label{sec:introduction}

% Introduce AI, IoT and Cloud continuum
Artificial Intelligence (AI) and the Internet of Things (IoT) paradigm are transforming the field of computing. AI-based applications are inherently compute-intensive and IoT introduces unprecedented decentralization making them communication-intensive as well. Cloud computing seems like a natural choice for these applications. The conventional cloud computing has evolved into today's edge (also known as \emph{cloudlets} or \emph{fog}) where computing occurs closer to the end devices that are typically mobile. Such a generic multi-tier cloud architecture, what we call the \emph{cloud continuum}, is shown in Figure~\ref{fig:cloud}.

% Relevance of time sensitivity in cloud resource allocation.
One of the foremost challenges in cloud resource allocation is the ability to satisfy the latency or deadline guarantees of an application. With the advent of 5G ultra-reliable low latency communication (uRLLC), time-sensitive applications such as telehealth, digital twins, and connected and autonomous cars, are expected to rely on the cloud continuum~\cite{ericsson}. For this reason, we expect to see an evolution of resource allocation techniques in the literature where the cloud continuum is modeled to handle time-sensitive applications, and hence these studies are the focus of this survey.

% Types of problem the papers address - offloading, provisioning, scheduling and taxonomy
Most works rely on a specific cloud model and define their own terminology. Therefore, we first define a generic cloud model and terminologies that encompass the surveyed literature. Existing works have majorly focused on three classes of problems: 1) The \emph{offloading decision} problem of whether to offload application computation from an end device to the edge and cloud or not. 2) The \emph{resource provisioning} problem of allocating the computation and/or communication resources to the applications. 3) The \emph{resource scheduling} problem of when to use the allocated computation and communication resources. The aim is to classify these works based on the type of problem they address, as well as the nature of the solution they propose (analytical or heuristic, centralized or decentralized, etc.). For time-sensitivity, we group the literature based on two objectives: response time minimization and satisfaction of hard deadlines.

% Paper choices - how do we select them
There are quite some works in the literature that consider time-sensitive applications. However, due to space limitations, it is not possible to cover all of them in this survey. We have therefore chosen papers based on the publication date (2013-2019) and the reputation of the venue (IEEE INFOCOMM, GLOBECOM, TPDS, TC, TCC, ICDCS). We suppose that studies earlier than 2013 are superseded by the later ones. Additionally, we filtered papers based on the quality of the proposed solution; those based on primitive heuristics or a simple application of optimization solvers are ignored. 
To the best of our knowledge, we are the first to survey resource allocation studies in the cloud continuum for time-sensitive applications.

% Organization of the paper
\paragraph{Organization.} The remainder of this article is organized as follows. Section~\ref{sec:architecture} describes the brief overview of the cloud model including the terminologies used in this paper. Section~\ref{sec:survey} reviews the existing cloud literature based on our taxonomy. We summarize the survey and identify some future research directions in Section~\ref{sec:conclusion}.

%% file: model.tex
\section{Multi-Tier Cloud Architecture}
\label{sec:architecture}
\begin{figure}[h!]
	\includegraphics[width=0.75\textwidth]{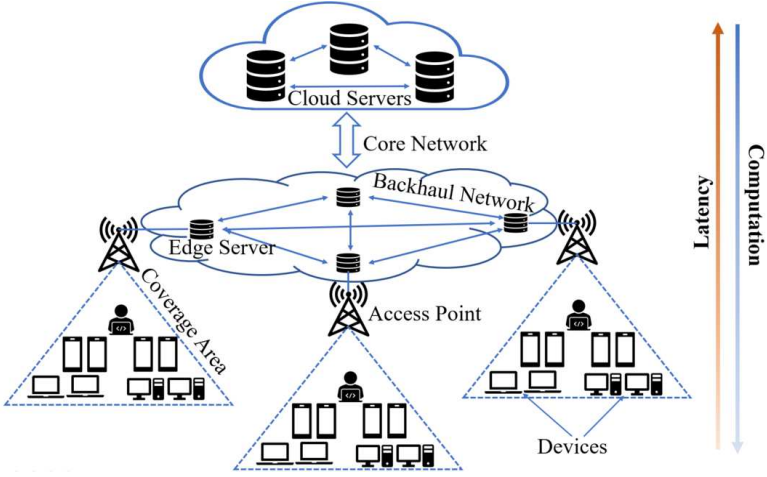}
	\caption{A multi-tier cloud architecture: both computation resources and access latency increase farther away from devices.}
	\label{fig:cloud}
	%\vspace{-0.3cm}
\end{figure}

% Brief introduction on how we classify papers, why this model is required, any terminologies used to describe the work
Existing literature models the cloud resources either as a collection of servers or as a set of servers interconnected by a backhaul network in a tiered architecture. Some studies consider the application workload as virtual machines (VM) with specific requirements (in terms of computation, storage, etc.) or abstract it using fractional requirements such as cycles/second (computation) or bits/second (communication). Hence, in order to classify this diverse literature there is a need to define a baseline cloud architecture model and terminologies (Figure \ref{fig:cloud} and Table~\ref{table:model}).

% Explain the cloud and edge server architecture and corresponding parameters.
The cloud servers, denoted by $N^C$, are the top-tier of the architecture with large amounts of resources. Each cloud server $n\in N^C$ has $C_{n,r}^C$ amount of type-$r$ resources. The cloud servers are connected to the edge servers with lower resource capacity by a high-speed core network. The amount of type-$r$ resource at the edge server location $n\in N^E$ is given by $C_{n,r}^E$. It is assumed that each edge server may have an access point through which the devices are connected to it. Each edge server has a bandwidth capacity for offloading workload tasks (generated by the devices). The servers are internally connected by a backhaul network. Generally, it is assumed that the core and the backhaul network have infinite bandwidth for data transmission.

The set of resources (processors, memory, storage) available at the cloud/edge is given by $R$. The set of virtual machines (VM's) of specific configuration or services at the server location $x$ is given by $M_x$ and the corresponding amount of type-$r$ resource required to host them is given by $H_{r,m}$, where $r\in R$, $m\in M_x$. Let $\mu_x$ denote the serving rate of the tasks at $x\in\{N^C\cup N^E\}$. 

The \emph{computation time} ($\delta_P$) of a task depends on the computation capacity of the server/device and any queuing delay. The computation capacity is computed either based on the computation speed (cycles per unit time) or the serving rate (tasks per unit time) of the server/device. Certain works assume servers have queues for storing tasks as their arrival rate may be higher than the computation capacity. This waiting time experienced by a task due to other tasks pending ahead of it is denoted as the \emph{queuing delay}.

% Explain the device model - device requests, minimum required parameters
The task (or device) $n\in N$ requests for a particular VM or service of type-$m$ for a specified duration (execution time) $T_{n,m}$. Each task is expected to be generated at a rate of $\lambda_n$. Each task may need to transfer data of size $S_n$ to the server and can have an offloading bandwidth of $B_n$. The task may specify whether it needs to be served within a deadline constraint $D_n$. There is a delay involved in sending the task data from the device to the servers or between servers. It is given by $d_{x,y}$, where $\{x,y\}\in\{N^C\cup N^E\cup N\}$. 

\emph{Communication time} ($\delta_T$) is the time required to transmit the data (\emph{i.e.,} $S_n/B_n$) from one entity to another including the communication delay ($d_{x,y}$). Several factors such as allocated bandwidth, interference, noise and distance play a role in determining this parameter. The elapsed time between a task's release and its completion is denoted as the \emph{response time}. This includes the computation time ($\delta_P$) and the communication time ($\delta_T$) for all entities on which the task executes. Some works also consider \emph{makespan}, which is the maximum response time among all the tasks.

\begin{table}[t]
	\centering
	\begin{tabular}{|c|l|}%{|c|p{6.3cm}|}
		\hline
		\textbf{Symbol}  & \multicolumn{1}{c|}{\textbf{Description}} \\ \hline
		\multicolumn{2}{|c|}{\textbf{Cloud Parameters}} \\ \hline
		$N^C$            & Set of cloud servers\\ \hline
		$C_{n,r}^C$      & Amount of type-$r$ resource available (capacity) at cloud server $n$\\ \hline
		\multicolumn{2}{|c|}{\textbf{Edge Parameters}}\\ \hline
		$N^E$            & Set of edge servers\\ \hline
		$C_{n,r}^E$      & Amount of type-$r$ resource available (capacity) at edge server $n$ \\ \hline
		$B_n^E$          & Bandwidth capacity of an edge server $n$ for task offloading\\ \hline
		\multicolumn{2}{|c|}{\textbf{Parameters common to Cloud and Edge}}\\ \hline
		$M_x$            & Set of VM's of specific configurations or services in $x$, where $x \in \{N^C \cup N^E\}$\\ \hline
		$R$            & Set of resources (Storage, Memory, CPUs)\\ \hline
		$H_{r,m}$       & Amount of type-$r$ resource required to host type-$m$ VM or service, where $r\in R$, $m\in M_x$\\ \hline
		$\mu_x$ & Serving rate of tasks in $x$, where $x \in \{N^C \cup N^E\}$\\ \hline
		\multicolumn{2}{|c|}{\textbf{Device/Task Parameter}}\\ \hline
		$N$            & Set of devices/tasks\\ \hline
		\multicolumn{2}{|c|}{\textbf{Task Parameters}}\\ \hline
		$\lambda_n$ & Arrival rate of task $n$\\ \hline
		$N_{n,m}$ & Number of type-$m$ VM's or services requested by task $n$\\ \hline
		$T_n$ & Duration of task $n$\\ \hline
		$D_n$ & Deadline constraint of task $n$\\ \hline
		$S_n$ & Data size of a task $n$\\ \hline
		$E_n$ & Constraint on edge server serving task $n$\\ \hline
		$B_n$ & Offloading bandwidth of task $n$\\ \hline
		\multicolumn{2}{|c|}{\textbf{Delay Parameter}}\\ \hline
		$d_{x,y}$ & Communication delay between entities $x$ and $y$ where $\{x,y\} \in \{N^C \cup N^E \cup N\}$\\ \hline
	\end{tabular}
	\caption{Model parameters. Note, $\mu_x$ and $T_n$ are mutually exclusive and either one of them can be used.}
	\label{table:model}
	%\vspace{-0.5cm}
\end{table}

%% file: classification_criteria.tex
%\vspace{-0.5cm}
\section{Literature Review}
\label{sec:survey}
In this section, we survey important resource allocation techniques that have been developed for the cloud continuum for time-sensitive applications. To classify this literature, we use the following taxonomy. 

\begin{enumerate}
	\item {\verb|Problem type.|} We consider two problem types; one based on the timing model and another based on the contention model. 
	\begin{enumerate}
	\item {\verb|Timing model.|} Studies that consider workload tasks with hard deadline requirements are classified under \textbf{deadline constrained} and presented in Section~\ref{sec:deadline}. The remaining works are categorized under \textbf{response time minimization}, including few studies that consider the makespan minimization problem, and presented in Section~\ref{sec:latency}.
	\item {\verb|Contention model.|} Depending on the contention model for the communication and/or the computation resources, the works are further classified as \textbf{no contention} (\emph{i.e.,} computation and communication resources are not shared between the tasks), \textbf{only communication contention} (\emph{i.e.,} tasks contend ONLY for offloading bandwidth $B_n$ and $\sum_n B_n$ is bounded by $B_n^E$), \textbf{only computation contention} (\emph{i.e.,} tasks contend ONLY for computation resources and in general, it is bounded by $C_{n,r}^C,C_{n,r}^E$ or $\mu_x$) and both \textbf{communication and computation contention}.
	\end{enumerate}
	\item {\verb|Solution type.|} We categorize the works based on the proposed solution type: \textbf{centralized} or \textbf{decentralized} algorithms. We further classify this based on the nature of solution.
	\begin{enumerate}
	\item {\verb|Nature of solution.|} Techniques that solve the problem or a relaxed variant of the problem either optimally or with an approximation bound are grouped under \textbf{analytical} solutions. The approximation bound could either be a \emph{constant} or depend on the task and server parameters (denoted as \emph{parameterized approximation bound}). The remaining works that propose heuristic techniques including meta-heuristic approaches are grouped under \textbf{heuristic} solutions.
	\end{enumerate}
\end{enumerate}

Table~\ref{tab:classification_table} shows the classification of literature based on the above taxonomy. We also identify the problem class (offloading, provisioning and scheduling) for each study in the same table. The literature review discussed in the subsequent sub-sections is based on the classification presented in this table.

%% file: survey.tex
\begin{table*}[]
	\centering
	\caption{Classification based on the problem type and the nature of solution proposed}
	\label{tab:classification_table}
	\resizebox{\textwidth}{!}{%
		\begin{tabular}{|c|c|c|c|c|c|}%{|L|L|L|C|C|J|L|L|K|K|}
			\hline
			& & \multicolumn{4}{c|}{\textbf{Solution Type}}\\ \cline{3-6} 
			& &	\multicolumn{2}{c|}{\textbf{Decentralized}}& \multicolumn{2}{c|}{\textbf{Centralized}}\\ \cline{3-6} 
			\multirow{-3}{*}{\textbf{Timing Model}}& \multirow{-3}{*}{\textbf{Contention Model}}& \textbf{Analytical}& \textbf{Heuristic}& \textbf{Analytical}& \textbf{Heuristic}\\ \hline
			& NO contention & & & \cite{Kao_2015}$^\S$ & \cite{Ding_2019}$^\S$\\ \cline{2-6}
			& Communication and computation & \cite{Cas_2019, Josilo_2019}$^\ast$ &	\cite{Duan_2014}$^\dagger$, \cite{Pang_2017}$^\ast$ & \cite{Gao_2019}$^\dagger$, \cite{Esh_2019}$^\ast$, \cite{Giroire_2019}$^\diamond$ & \cite{Heydari_2019}$^\S$,  \cite{Chen_M_2017,Saleem_2018}$^\ast$\\ \cline{2-6}
			\multirow{-2}{*}{\textbf{\begin{tabular}[m]{@{}c@{}}Response Time\\ Minimization\end{tabular}}}  & ONLY computation &  \cite{Xiao_2017}$^\dagger$, \cite{Tan_2017}$^\diamond$ & \cite{Abouaomar_2018, Jin_2017}$^\dagger$ & \begin{tabular}[m]{@{}c@{}} \cite{Cao_2014, Di_2013, Jin_2017, Xu_2016, Zha_2014}$^\dagger$, \cite{Ouyang_2019, Ren_2017}$^\ast$\\ \cite{Chen_2017, Shu_2017, Ton_2016, Zhang_2017}$^\diamond$ \end{tabular} & \begin{tabular}[m]{@{}c@{}} \cite{Tarplee_2016, Zeng_2016}$^\ddagger$, \cite{Yaqub_2018}$^\ast$, \\ \cite{Shu_2017, Ton_2016, Zhang_2017}$^\diamond$, \cite{Han_2019, Yang_2015}$^\circ$ \end{tabular} \\ \cline{2-6}
			& ONLY communication & \cite{Chen_2016}$^\ast$ & \cite{Liu_2018a}$^\S$ & \cite{Mao_2016}$^\ast$ & \\ \cline{2-6}
			\noalign{\vskip\doublerulesep\vskip-\arrayrulewidth} \cline{1-6}
			& NO contention & & \cite{Zhang_2019}$^\dagger$ & \cite{ChenL_2019}$^\dagger$, \cite{Guo_2017, Kao_2014, Millnert_2019}$^\diamond$ &  \\ \cline{2-6}
			& Communication and computation & & \cite{Millnert_2018}$^\dagger$, \cite{Vu_2019}$^\ast$ & \cite{Cziva_2018}$^\dagger$, \cite{Vu_2018}$^\ast$, \cite{Meng_2019, Zheng_2016}$^\diamond$ & \\ \cline{2-6}
			\multirow{-2}{*}{\textbf{\begin{tabular}[m]{@{}c@{}}Deadline\\ Constrained\end{tabular}}}	& ONLY computation & & \cite{Chang_2017}$^\S$, \cite{Zhu_2015}$^\diamond$ & \begin{tabular}[m]{@{}c@{}} \cite{Chang_2017}$^\S$, \cite{Chen_2019, Gu_2015, Liu_2018, Mei_2015}$^\dagger$, \\ \cite{Yu_2015}$^\ddagger$, \cite{Dai_2018,Du_2018}$^\ast$, \cite{Yin_2017}$^\diamond$ \end{tabular} & \begin{tabular}[m]{@{}c@{}} \cite{Ma_2019, Wei_2018}$^\dagger$, \cite{Hu_2018, Wu_2017}$^\ddagger$ \\ \cite{Begam_2018, Cai_2019, Cal_2014, Fan_2017, Rod_2014, Sundar_2018, Wang_2015}$^\diamond$ \end{tabular} \\ \cline{2-6}
            & ONLY commmunication & & & \cite{Guo_2016}$^\S$, \cite{Sun_C_2017}$^\dagger$, \cite{Tong_2016}$^\ddagger$, \cite{Nyugenl_2017, Yu_2018}$^\ast$, \cite{Gao_2017}$^\circ$ & \cite{Nyugenl_2017}$^\ast$, \cite{Tong_2016}$^\ddagger$ \\ \hline
		\end{tabular}%
    }
    \begin{tablenotes}
    	\small
    	\item  $\S \gets$ Offloading; $\dagger \gets$ Provisioning; $\ddagger \gets$ Scheduling; $\ast \gets$ Offloading and provisioning; $\diamond \gets$ Provisioning and scheduling; $\circ \gets$ Offloading, provisioning and scheduling.
    	\end{tablenotes}
\end{table*}

\subsection{Response Time Minimization}
\label{sec:latency}
Many studies aim to minimize the latency experienced by tasks under various constraints. The most common timing-related objective found in these studies is that of task response time minimization. These include minimizing the average task response times (\emph{i.e., }$\min \sum_{N} (\delta_P + \delta_T$)) or minimizing the overall makespan (\emph{i.e., }$\min \max_{\forall N} (\delta_P + \delta_T)$). In this section, we review the literature that consider these two problems and categorize them based on their respective contention model. 

\subsubsection{No contention.}
Works in this category mainly focus on the task offloading problem on single-tier architectures with optimization objectives such as minimizing task response times~\cite{Kao_2015} and device energy~\cite{Ding_2019}.

%% Offloading
Kao and Krishnamachari \cite{Kao_2015} model the workload as a Directed Acyclic Graph (DAG) where vertices represent tasks and edges represent data dependencies among them. Using dynamic programming the DAG is split into multiple trees and the response-time of each tree is optimized using time quantization, as in~\cite{Kao_2014}. 
They present a Fully Polynomial Time Approximation Scheme (FPTAS) with an approximation factor of $(1+\epsilon)$, where $\epsilon\in[0,1]$ is chosen by users to reach a trade-off between optimality and algorithm runtime. 
Ding \emph{et al.}~\cite{Ding_2019} formulate the problem as a Mixed Integer Non-Linear Problem (MINLP) with a fixed offloading bandwidth for tasks. They reduced it to a Quadratically Constrained Quadratic Programming (QCQP) problem and apply semi-definite relaxation (SDR) to obtain optimal offloading decisions using optimization solvers.

\subsubsection{Communication and computation contention.}
Studies in this category mainly focus on the task offloading and server provisioning problems with optimization objectives such as minimizing task response times~\cite{Cas_2019, Esh_2019, Gao_2019, Heydari_2019, Josilo_2019}, makespan~\cite{Duan_2014, Giroire_2019, Pang_2017}, device energy~\cite{Esh_2019, Saleem_2018}, server usage costs and communication overhead~\cite{Chen_M_2017, Duan_2014}.

\seimav{Heydari \emph{et al.}~\cite{Heydari_2019} consider the task offloading problem on a single-tier architecture. They formulate the problem as a Markov Decision Process and propose an actor-critic based reinforcement learning heuristic to learn the offloading decisions.}

\ebv{Some studies consider the server provisioning problem on single-tier~\cite{Duan_2014} and multi-tier architectures~\cite{Gao_2019}. Gao \emph{et al.}~\cite{Gao_2019} formulate it as a Pure Integer Non-Linear Programming (PINLP) problem as well as a sub-divided Integer Non-Linear Programming (INLP) problem. They propose a lazy switch algorithm to control the task migration frequency between servers and use a solver for the INLP iteratively, providing a parameterized performance approximation bound.} Duan \emph{et al.}~\cite{Duan_2014} model tasks as a DAG and propose a decentralized online algorithm based on cooperative sequential games for the problem of allocating processors across servers to each DAG node, where the allocated bandwidth capacity is also proportional to the number of allocated processors.
	
\arv{Some studies consider the combined task offloading and server provisioning problem on single-tier~\cite{Cas_2019, Josilo_2019, Pang_2017, Saleem_2018} and multi-tier architectures~\cite{Esh_2019, Chen_M_2017}. Modeling task response times generically using server-specific utility functions,~\cite{Cas_2019} presents a decentralized max-consensus based greedy algorithm for the problem with a constant approximation bound of $(1-1/e)$ and shows polynomial-time convergence under some conditions on the utility function. On the other hand, Jo\v{s}ilo \emph{et al.}~\cite{Josilo_2019} model the problem in a decentralized game-theoretic framework, and derive a policy with guaranteed convergence to a Nash equilibrium using Stackelberg games with a constant approximation bound of $(3+\sqrt{5})/2$}. Pang \emph{et al.}~\cite{Pang_2017} propose a heuristic using dynamic programming where the servers provision resources in proportion to the amount of resources requested in a decentralized manner by exchanging information on the tasks. \ebv{Saleem \emph{et al.}~\cite{Saleem_2018} formulate an MINLP optimization problem with energy constraints and propose a greedy heuristic to allocate communication resources based on tasks' offloading bandwidth.} \nsv{Eshraghi and Liang~\cite{Esh_2019} formulate a non-convex mixed-integer problem which is further reduced to a convex form with binary relaxation. They provide an optimal solution using a geometric programming that is iteratively applied on each processor of a multi-processor server.} \ebv{Chen \emph{et al.}~\cite{Chen_M_2017} formulate it as a QCQP problem and propose a heuristic combining SDR, alternating optimization and sequential tuning, and provide a lower bound on server usage cost}.

\arv{Giroire \emph{et al.}~\cite{Giroire_2019} consider the joint server provisioning and task scheduling problem on single-tier architectures. They model tasks as a DAG and propose a greedy list scheduling algorithm based on communication overhead that is optimal for tasks with constant response times and bounded bandwidth capacity. Further, they extend the solution with parameterized approximation algorithms using k-balanced (k-servers) partitioning for tasks with unbounded bandwidth capacity.}

\subsubsection{Only computation contention.}
In this category, studies mainly focus on the server provisioning and task scheduling problems with optimization objectives such as minimizing task response times~\cite{Abouaomar_2018, Cao_2014, Chen_2017, Di_2013, Ren_2017, Shu_2017, Tan_2017, Ton_2016, Xiao_2017, Xu_2016, Yang_2015, Zhang_2017}, makespan~\cite{Zeng_2016}, device energy~\cite{Yaqub_2018}, server energy~\cite{Jin_2017, Tarplee_2016, Zha_2014}, server usage costs~\cite{ Han_2019, Ouyang_2019} and communication overhead~\cite{Han_2019}.

% Provisioning
Some studies focus on VM and server provisioning problems on single-tier~\cite{Abouaomar_2018, Cao_2014, Di_2013, Jin_2017, Xu_2016, Zha_2014} and multi-tier~\cite{Xiao_2017} architectures. Abouaomar \emph{et al.}~\cite{Abouaomar_2018} propose a matching game-based heuristic solution to identify servers for offloading using a decentralized deferred acceptance algorithm. 
Cao \emph{et al.}~\cite{Cao_2014} model the response time using an M/M/m queuing model where m is the number of servers, and solve the optimization problem using Lagrange multipliers and bisection methods for optimal server speed and workload arrival. 
\nsv{Di and Wang~\cite{Di_2013} model the response time as a ratio of the task workload over its allocated resources, both abstracted with input parameters. The optimal resource allocation for each task is then determined using the Karush-Kuhn-Tucker (KKT) conditions in polynomial time.} Modeling the response time as a function of the queuing delay on servers,~\cite{Zha_2014} proposes a centralized online algorithm with a parameterized approximation bound, using integer relaxation to a linear programming (LP) problem and first-fit strategy to subsequently satisfy the integrality constraints. On the other hand,~\cite{Jin_2017} models the response time as a function of the number of co-allocated VMs, and proposes a centralized online greedy algorithm with a parameterized approximation bound by sorting the VMs based on their arrival order. It also proposes a decentralized heuristic extension to this algorithm where each server performs a cost-benefit analysis comparing the cost of provisioning a VM alone to the incremental cost of provisioning that VM given the current provisions. Considering a single-tier architecture made up of interconnected access points,~\cite{Xu_2016} proposes a graph representation method to solve the problem using capacitated k-median problem and derives parameterized approximation bounds. 
While considering multi-tier architectures, Xiao and Krunz~\cite{Xiao_2017} propose a decentralized strategy using Lagrange decomposition to transform the global provisioning problem into server-specific convex optimization problems. They also show that the proposed strategy converges to the global optimum at a rate inversely proportional to the number of iterations.

% Scheduling
A few studies focus on the task scheduling problem on single-tier~\cite{Tarplee_2016} and multi-tier~\cite{Zeng_2016} architectures. Tarplee \emph{et al.}~\cite{Tarplee_2016} formulate the problem as an ILP and solve using a relaxation method, where they assume tasks can be decomposed in chunks of arbitrary size to be run in parallel. They propose a heuristic solution based on the Convex Fill algorithm.
Whereas,~\cite{Zeng_2016} uses an M/M/1 queuing model and formulates the problem as an MINLP. It decomposes the problem into sub-problems and proposes a heuristic solution to solve each sub-problem sequentially using LP relaxation.

Some studies consider the joint task offloading and server provisioning problem on single-tier~\cite{Ren_2017, Yaqub_2018} and multi-tier~\cite{Ouyang_2019} architectures. Ren \emph{et al.}~\cite{Ren_2017} formulate response-time minimization as a piece-wise convex function to determine the optimal proportion of each task to be executed on the device and the server with a fixed offloading bandwidth per task. For the special case of limited device computation capacity, the optimal length of Time Division Multiple Access (TDMA) slots is also computed for each device. Yaqub and Sorour~\cite{Yaqub_2018} present a priority-based heuristic and bisection method for offloading decisions on neighboring devices and servers, respectively. They propose a heuristic solution for the provisioning problem using the Lagrangian method. 
Ouyang \emph{et al.}~\cite{Ouyang_2019} propose an offline solution using the shortest path algorithm for DAG tasks. It also presents an online learning algorithm for provisioning using multi-arm bandit with a parameterized regret bound.

% Provisioning and Scheduling
Some studies focus on both server provisioning and task scheduling problems on single-tier~\cite{Chen_2017, Shu_2017, Zhang_2017} and multi-tier~\cite{Tan_2017, Ton_2016} architectures. 
Considering max-min fairness, which maximizes the minimum resource allocation across tasks sharing servers, Chen \emph{et al.}~\cite{Chen_2017} reduce the optimization problem to an LP for a single task case and find the optimal solution. For multiple tasks, they iterate the procedure to ensure max-min fairness.  
Considering DAG tasks, Shu \emph{et al.}~\cite{Shu_2017} propose an FPTAS for the makespan minimization problem through a reduction to the constrained shortest path problem for single-resource VMs. For the more general case of multi-resource VMs, they propose a greedy heuristic based on critical paths and binary search. 
Zhang \emph{et al.}~\cite{Zhang_2017} present a priority-based weighted algorithm for provisioning with a constant approximation bound of $2$ in terms of the number of  servers and a heuristic scheduling algorithm based on the Karmarkar-Karp differencing algorithm. 
Tong \emph{et al.}~\cite{Ton_2016} consider fractional resource allocations with a fixed offloading bandwidth per task. For the special case of one server per tier of the architecture, they present optimal centralized solutions using convex optimization and branch-and-bound methods, whereas, for the more general problem, they present a solution based on simulated annealing. Tan \emph{et al.}~\cite{Tan_2017} propose a decentralized solution by selecting the server with the least increase in response time and schedule using the shortest remaining computation time first policy. They prove this to be $\mathcal{O}(1/\epsilon)$-competitive (in terms of response time) with a corresponding constant approximation bound of $1+\epsilon$ on the speed of servers.

% Offloading , Provisioning and Scheduling
A few studies consider the joint problem of task offloading, server provisioning and task scheduling on single-tier~\cite{Han_2019, Yang_2015} architectures. Considering DAG tasks, Han \emph{et al.}~\cite{Han_2019} present a priority-based heuristic solution where tasks are sorted based on total average computation and communication time. Considering sequential tasks with a constraint on the number of tasks allocated per server, Yang \emph{et al.}~\cite{Yang_2015} propose a greedy heuristic in which tasks are first offloaded to the server without any resource constraint and later to meet the constraint some tasks are moved back based on a reward function.

\subsubsection{Only communication contention.}
In this category, works focus on task offloading and bandwidth provisioning problem with optimization objectives such as minimizing device energy~\cite{Chen_2016, Liu_2018a, Mao_2016} and server energy~\cite{Chen_2016, Liu_2018a}.

Chen \emph{et al.}~\cite{Chen_2016} model the bandwidth as a function of the interference among tasks in the wireless network. They model the problem in a decentralized game-theoretic framework to minimize both task response time and makespan on single-tier architectures. They derive a policy using potential games with finite improvement property with guaranteed convergence to a Nash equilibrium and a parameterized approximation bound.
Mao \emph{et al.}~\cite{Mao_2016} formulate it as a stochastic optimization problem for multi-tier architectures. They propose a Lyapunov optimization-based algorithm and use the Lagrangian method and KKT conditions to determine the optimal device power and offloading bandwidth.
On the other hand, Liu \emph{et al.}~\cite{Liu_2018a} consider only the task offloading problem on a single-tier architecture. They derive a heuristic policy using population games, where player strategies are modeled using a Markov evolutionary process.

\subsection{Deadline Constrained}
\label{sec:deadline}
Most time-critical tasks request for resources with a notion of a deadline. In this section, we assume that the deadline defines a requirement on the task's response time which includes both computation and communication times, unless specified otherwise. We present all studies that consider workload tasks with such deadlines, irrespective of the optimization objective they address.

\vspace{-0.2cm}
\subsubsection{No contention.}

In this category, works mainly focus on the server provisioning and/or task scheduling problems with optimization objectives such as minimizing task response times~\cite{ChenL_2019, Millnert_2019}, device energy~\cite{Guo_2017, Kao_2014} and task deadline misses~\cite{Zhang_2019}.

A few studies only consider the server provisioning~\cite{ChenL_2019, Zhang_2019} problem on single-tier~\cite{ChenL_2019} and multi-tier architectures~\cite{Zhang_2019}. Chen \emph{et al.}~\cite{ChenL_2019} additionally consider a greedy task replication strategy for fault tolerance and propose a multi-arm bandit learning algorithm with a parameterized approximation bound for sub-modular marginal reward functions (reward is based on a probabilistic prediction of task completion times). Zhang \emph{et al.}~\cite{Zhang_2019} use a singleton weighted congestion game based heuristic to arrive at a consensus on task allocation at the lower tier. They also use a stochastic Lyapunov optimization-based greedy heuristic to estimate task response times and decide whether to admit the task or to provision it on another server at the higher tier. 

%% provisioning and scheduling
Some studies consider the server provisioning and task scheduling problems on single-tier~\cite{Kao_2014, Guo_2017} and multi-tier~\cite{Millnert_2019} architectures. 
Considering a set of task flows allocated on a resource graph where each flow is a sequence of sub-tasks with an end-to-end deadline, Millnert \emph{et al.}~\cite{Millnert_2019} present a centralized analytical technique for dynamic adjustments to the response times experienced by tasks. They propose protocols that use an upper bound on the rate of change of response times which would ensure the satisfaction of all end-to-end deadlines. They present protocols for dynamically changing task flows as well as resource graphs. On the other hand, considering tasks modeled as a collection of trees with end-to-end deadlines and fixed offloading bandwidth,~\cite{Kao_2014} presents a centralized dynamic programming based polynomial-time solution using time quantization, and an exponential-time extension for tasks with probabilistic computation times.
Guo \emph{et al.}~\cite{Guo_2017} formulate the convex optimization problem as a three-stage flow-shop scheduling problem by separately considering the offloading, constant execution and downloading duration of each task. They solve the problem optimally when the minimum offloading duration is larger than the maximum execution duration of all tasks using KKT conditions and bisection search method.

\subsubsection{Communication and computation contention.}
Studies in this category mainly focus on server provisioning and task scheduling problems with optimization objectives such as minimizing VM delays~\cite{Cziva_2018, Millnert_2018}, task deadline misses~\cite{Meng_2019}, device energy~\cite{Vu_2018, Vu_2019} and maximizing task utility~\cite{Zheng_2016}.

Some studies consider the problem of server provisioning for single-tier~\cite{Cziva_2018} and multi-tier~\cite{Millnert_2018} architectures with re-provisioning for changes in the device coverage area. Cziva \emph{et al.}~\cite{Cziva_2018} model the resources of the servers with bounded bandwidth capacity and communication delay, and propose a technique using Optimal Stopping theory. Millnert \emph{et al.}~\cite{Millnert_2018} consider task flows pre-allocated on a resource graph with end-to-end deadlines as in~\cite{Millnert_2019}, and present a decentralized heuristic solution through deadline decomposition based on control theoretic and optimization frameworks to reduce VM creation delays.

A few studies consider the task offloading and server provisioning problem on multi-tier~\cite{Vu_2018, Vu_2019} architectures. 
Vu \emph{et al.}~\cite{Vu_2018} formulate the problem as an MINLP, and use integer relaxation and branch and bound algorithm to find an optimal solution and prune the search space. They extend this in~\cite{Vu_2019} with additional parameters such as offloading and downloading bandwidth. They propose a decentralized heuristic algorithm through decomposition using the bender's cuts.

A few works focus on the task provisioning and scheduling problems on single-tier~\cite{Zheng_2016} and multi-tier architectures~\cite{Meng_2019}. \nsv{Zheng and Shroff~\cite{Zheng_2016} propose an online algorithm for stochastic tasks in the continuous and discrete-time domain with a competitive ratio of 2 and 1.8, respectively.} On the other hand,~\cite{Meng_2019} proposes an online heuristic based on the largest computation time to reduce the number of deadline misses and derives a parameterized competitive ratio on the makespan.

\input{table}

\subsubsection{Only computation contention.}
In this category, studies mainly focus on the server provisioning and task scheduling problems with optimization objectives such as minimizing the task response times~\cite{Dai_2018,Du_2018,Liu_2018,Wang_2015}, device energy~\cite{Chang_2017,Du_2018}, server energy~\cite{Chen_2019, Gu_2015, Yu_2015}, server usage costs~\cite{Cai_2019, Cal_2014, Fan_2017, Ma_2019, Mei_2015, Rod_2014, Sundar_2018, Wu_2017, Yin_2017}, peak resource utilization on servers~\cite{Hu_2018, Wei_2018}, task deadline misses~\cite{Begam_2018, Mei_2015, Zhu_2015} and communication overhead~\cite{Sundar_2018}.

%Normal tasks, both provisioning and scheduling
Some studies focus on both server provisioning and task scheduling problems on single-tier~\cite{Begam_2018, Wang_2015, Yin_2017, Zhu_2015} or multi-tier~\cite{Fan_2017,Sundar_2018} architectures. Considering a variety of different objectives, they propose heuristic solutions using techniques such as prioritization based on task parameters with best-fit provisioning~\cite{Begam_2018}, agent-based decentralized bidding between tasks and server VMs based on task parameters~\cite{Zhu_2015}, as early as possible scheduling with load balancing~\cite{Wang_2015}, ant colony optimization with a response time dependent utility function~\cite{Fan_2017}, and a discretization strategy that combines the provisioning results of a convex optimization solver with greedy deadline-driven scheduling~\cite{Sundar_2018}. Wang \emph{et al.}~\cite{Wang_2015} also consider fault tolerance using backup tasks that are executed as late as possible with their allocations being reclaimed when not required. Yin \emph{et al.}~\cite{Yin_2017} formulate it as an LP relaxation and solve using dual decomposition with an online algorithm that has a parameterized competitive ratio in terms of resource capacity augmentation when compared to an optimal offline algorithm.

%Normal tasks, only provisioning
Some studies only focus on the server provisioning problem on single-tier~\cite{Chen_2019, Gu_2015, Liu_2018, Mei_2015, Wei_2018} and multi-tier architectures~\cite{Ma_2019}. Again considering a variety of different objectives, they either present analytical~\cite{Chen_2019, Gu_2015, Liu_2018, Mei_2015} or heuristic~\cite{Wei_2018, Ma_2019} solutions. Chen \emph{et al.}~\cite{Chen_2019} consider a demand-response setting that enforces a maximum peak power for each server. They present an online solution with a parameterized approximation bound using Vickrey-Clark-Groves (VCG) auctions and also consider the trade-off between switching costs and energy loss for server activations and deactivations. Gu \emph{et al.}~\cite{Gu_2015} and Liu \emph{et al.}~\cite{Liu_2018} formulate MINLP problems and optimally solve relaxed duals using either block coordinate descent method~\cite{Liu_2018} or Lagrangian with a dynamic voltage and frequency scaling strategy~\cite{Gu_2015}. Considering a $M/M/m$ queuing model, Mei \emph{et al.}~\cite{Mei_2015} optimally solve the problem using bisection method assuming the number of servers $m$ and the speed of each server are continuous variables. Then, they recover integer values for these variables with the least server usage costs. 
Considering a bound on VM allocation delay, Wei \emph{et al.}~\cite{Wei_2018} propose an online greedy heuristic strategy based on balancing the remaining resource capacities across servers with future workload predictions modeled as a Markov chain that uses moving averages~\cite{Wei_2018}. Ma \emph{et al.}~\cite{Ma_2019} model the costs separately for on-demand and reserved resource provisioning on servers, and present heuristics based on gradient descent, bisection method and piece-wise convex optimization to provision reserved, on-demand and both the resources, respectively. 

%Normal tasks, miscellaneous
A few studies consider either the task offloading problem~\cite{Chang_2017}, the joint task offloading and server provisioning problem~\cite{Dai_2018, Du_2018} or the task scheduling problem~\cite{Yu_2015} on single-tier~\cite{Chang_2017,Dai_2018,Yu_2015} and multi-tier~\cite{Du_2018} architectures. Chang \emph{et al.}~\cite{Chang_2017} use queuing theory and show that the presented centralized solution is guaranteed to converge to the optimal value because the objective function is quasi-convex. They also propose a decentralized heuristic that uses Lagrange decomposition and transforms the global problem into device-specific relaxed convex optimization problems. Dai \emph{et al.}~\cite{Dai_2018} consider fixed offloading bandwidth for tasks and iteratively solve the joint problem as an MINLP, where the offloading problem is relaxed to a real-valued NLP and solved using bipartite graph-based rounding method with a parameterized approximation bound, and the provisioning problem is solved optimally using Lagrangian multipliers with a gradient descent method. Du \emph{et al.}~\cite{Du_2018} minimize the weighted sum of task response time and device energy with a fixed offloading bandwidth for tasks. They formulate it as a QCQP and reduce it to a convex problem using SDR and use the bisection method to determine the offloading decisions. They present a sub-optimal power and offloading bandwidth allocation algorithm using Lagrange multipliers. Finally, Yu \emph{et al.}~\cite{Yu_2015} model energy costs as battery losses. They transform the problem into a queue stability problem using the framework of Lyapunov optimization and present an algorithm for task and battery scheduling with a parameterized approximation bound, where admission control is performed based on the available server capacity. 

%DAGs 
Some studies model the workload with a DAG and end-to-end deadline constraint on the DAG~\cite{Cai_2019,Cal_2014,Hu_2018,Rod_2014,Wu_2017}, where the nodes are tasks and the edges are precedence constraints among tasks. 
Focusing on both server provisioning and DAG scheduling problems on single-tier architectures, studies present heuristic solutions using particle swarm optimization~\cite{Rod_2014} and greedy deadline decomposition and scheduling strategies based on slowest-cheapest VMs and earliest ready tasks with fixed offloading bandwidth~\cite{Cai_2019}. Extensions to handle variations in the computation and communication times using task replication and critical path detection have also been proposed~\cite{Cal_2014}. Note, in these studies, although the scheduling problem uses a contention model for computation time, the provisioning problem is modeled without contention by allowing for an arbitrary number of VM instantiations. Other studies only consider the DAG scheduling problem on single-tier architectures, and propose deadline decomposition-based heuristic solutions~\cite{Hu_2018,Wu_2017}. Hu \emph{et al.}~\cite{Hu_2018} use LP by converting the DAG to a set of independent task groups with deadlines decomposed in proportion to the number and computation time of tasks in each group. Whereas, Wu \emph{et al.}~\cite{Wu_2017} use probabilistic list scheduling with tasks ordered using ant colony optimization and deadlines decomposed based on critical paths.

\subsubsection{Only communication contention.}
Studies in this category mainly focus on task offloading and server/bandwidth provisioning problems with optimization objectives such as minimizing server energy~\cite{Sun_C_2017}, device energy~\cite{Nyugenl_2017}, server usage costs and communication overhead~\cite{Gao_2017, Guo_2016, Yu_2018}.

Some studies consider the task offloading problem~\cite{Guo_2016}, bandwidth provisioning problem~\cite{Sun_C_2017} and joint task offloading and server provisioning problem~\cite{Yu_2018} on single-tier architectures. 
Guo \emph{et al.}~\cite{Guo_2016} model the tasks as DAGs and formulate the problem as a non-convex problem. They relax it and optimally solve its dual problem using Lagrangian multipliers and sub-gradient method. 
Considering a bound on queuing delay, Sun \emph{et al.}~\cite{Sun_C_2017} derive a probability function for deadline misses and use interior point method to find the optimal solution. 
Yu \emph{et al.}~\cite{Yu_2018} formulate the problem as a multi-commodity max-flow problem and propose an FPTAS assuming tasks can be arbitrarily parallelized. They also propose a randomized algorithm with a parameterized approximation bound for tasks that are not parallelizable. Considering multi-tier architectures, Nguyen \emph{et al.}~\cite{Nyugenl_2017} formulate the joint problem as a min-max INLP and use the bisection search method to compute the optimal device frequency and wireless channel assignment. They also present a low-complexity heuristic solution using decoupled ILP based optimization.

Tong and Gao~\cite{Tong_2016} only consider the wireless network scheduling problem on single-tier architectures. They propose a dynamic programming solution, for a burst of transmissions, by computing the optimal delay in task communications. 
Gao \emph{et al.}~\cite{Gao_2017} focus on the joint task offloading, server provisioning and task scheduling problem on a multi-tier architecture with a bound on communication delay. They propose a greedy offline algorithm based on a task-specific utility function and an opportunistic online algorithm in which tasks offload in the first convenient slot they find, both with an approximation bound of $2$.

%% file: table.tex
\begin{table*}[h!]
		\centering
		\caption{Literature classification based on timing related model parameters}
		\label{tab:classification_table2}
		\resizebox{\textwidth}{!}{%
		\begin{tabular}{|L|J|L|C|C|J|L|L|K|K|}
		\hline
		\multirow{2}{*}{\textbf{\begin{tabular}[m]{@{}c@{}}Cloud\\ Architecture\end{tabular}}} & \multicolumn{3}{c|}{\textbf{Server Parameters}} & \multicolumn{4}{c|}{\textbf{Task Parameters}} & \multicolumn{2}{c|}{\textbf{Delay Parameters}}  \\ \cline{2-10}
		& \textbf{\begin{tabular}[c]{@{}c@{}}Computation\\ capacity\end{tabular}} & \textbf{\begin{tabular}[c]{@{}c@{}}Bandwidth\\ capacity\end{tabular}} & \textbf{\begin{tabular}[c]{@{}c@{}}Serving\\ rate\end{tabular}} & \textbf{\begin{tabular}[c]{@{}c@{}}Arrival\\ rate\end{tabular}} & \textbf{\begin{tabular}[c]{@{}c@{}}Duration /\\ Execution time\end{tabular}} & \textbf{\begin{tabular}[c]{@{}c@{}}Deadline\\ constraint\end{tabular}} & \textbf{\begin{tabular}[c]{@{}c@{}}Offloading\\ bandwidth\end{tabular}} & \textbf{Device-Server} & \textbf{Server-Server} \\ \hline
		\textbf{Single-tier}& \cite{Abouaomar_2018, Begam_2018, Cai_2019, Cas_2019, Chen_2017, Chen_2019, Cziva_2018, Dai_2018, Di_2013, Duan_2014, Giroire_2019, Gu_2015, Han_2019, Heydari_2019, Hu_2018, Jin_2017, Josilo_2019, Liu_2018, Liu_2018a, Pang_2017, Ren_2017, Rod_2014, Saleem_2018, Shu_2017, Tarplee_2016, Wang_2015, Wei_2018, Wu_2017, Xu_2016, Yang_2015, Yaqub_2018, Yin_2017, Yu_2015, Zha_2014, Zhang_2017, Zheng_2016, Zhu_2015} & \cite{Cai_2019, Cas_2019, Chen_2016, Chen_2017, Cziva_2018, Dai_2018, Ding_2019, Duan_2014, Giroire_2019, Guo_2016, Josilo_2019, Kao_2014, Heydari_2019, Liu_2018a, Mao_2016, Pang_2017, Saleem_2018, Sun_C_2017, Tong_2016, Yu_2018, Zheng_2016} & \cite{Cao_2014, Chang_2017, Mei_2015, Sun_C_2017, Wei_2018, Yu_2015, Zha_2014, Zhang_2017} & \cite{Cao_2014, Chang_2017, Gu_2015, Mei_2015, Sun_C_2017, Wei_2018, Yu_2015, Zha_2014, Zhang_2017} & \cite{Abouaomar_2018, Begam_2018, Cai_2019, Cal_2014, Cao_2014, Cas_2019, Chang_2017, Chen_2016, Chen_2017, Chen_2019, ChenL_2019, Dai_2018, Di_2013, Ding_2019, Duan_2014, Giroire_2019, Guo_2016, Guo_2017, Han_2019, Heydari_2019, Hu_2018, Jin_2017, Josilo_2019, Kao_2014, Kao_2015, Liu_2018, Liu_2018a, Mao_2016, Pang_2017, Ren_2017, Saleem_2018, Shu_2017, Tarplee_2016, Wu_2017, Xu_2016, Yang_2015, Yaqub_2018, Yin_2017, Yu_2015, Zhu_2015} & \cite{Begam_2018, Cai_2019, Cal_2014, Chang_2017, Chen_2019, ChenL_2019, Cziva_2018, Dai_2018, Gu_2015, Guo_2016, Guo_2017, Hu_2018, Kao_2014, Liu_2018, Mei_2015, Rod_2014, Sun_C_2017, Tong_2016, Wang_2015, Wei_2018, Wu_2017, Yin_2017, Yu_2015, Yu_2018, Zheng_2016, Zhu_2015} & \cite{Abouaomar_2018, Cai_2019, Cas_2019, Chang_2017, Chen_2016, ChenL_2019, Cziva_2018, Dai_2018, Ding_2019, Duan_2014, Giroire_2019, Guo_2016, Guo_2017, Han_2019, Heydari_2019, Josilo_2019, Kao_2014, Liu_2018, Liu_2018a, Mao_2016, Pang_2017, Ren_2017, Saleem_2018, Sun_C_2017, Tong_2016, Wu_2017, Yaqub_2018, Yu_2018} & \cite{Chen_2017, ChenL_2019, Cziva_2018, Ding_2019, Guo_2016, Guo_2017, Heydari_2019, Josilo_2019, Kao_2015, Liu_2018, Pang_2017, Ren_2017, Saleem_2018, Sun_C_2017, Tong_2016, Yang_2015} & \cite{Cas_2019, Chen_2019, ChenL_2019, Di_2013, Kao_2015, Wei_2018, Xu_2016, Yu_2018} \\ \hline
		\textbf{Multi-tier}& \cite{Chen_M_2017, Du_2018, Esh_2019, Fan_2017, Gao_2019, Meng_2019, Millnert_2018, Ouyang_2019, Sundar_2018, Tan_2017, Ton_2016, Vu_2018, Vu_2019} & \cite{Chen_M_2017, Du_2018, Esh_2019, Fan_2017, Gao_2017, Gao_2019, Meng_2019, Millnert_2018, Nyugenl_2017, Ton_2016, Vu_2018, Vu_2019} & \cite{Gao_2019, Ma_2019, Millnert_2018, Xiao_2017, Zeng_2016} & \cite{Gao_2019, Ma_2019, Millnert_2018, Xiao_2017, Zeng_2016} & \cite{Chen_M_2017, Du_2018, Esh_2019, Fan_2017, Meng_2019, Nyugenl_2017, Ouyang_2019, Sundar_2018, Tan_2017, Ton_2016, Vu_2019, Vu_2018, Zhang_2019} & \cite{Du_2018, Fan_2017, Gao_2017, Ma_2019, Meng_2019, Millnert_2018, Millnert_2019, Nyugenl_2017, Sundar_2018, Vu_2018, Vu_2019, Zhang_2019} & \cite{Chen_M_2017, Du_2018, Esh_2019, Gao_2017, Gao_2019, Meng_2019, Nyugenl_2017, Ton_2016, Vu_2018, Vu_2019} & \cite{Chen_M_2017, Du_2018, Esh_2019, Fan_2017, Gao_2019, Ma_2019, Meng_2019, Nyugenl_2017, Ouyang_2019, Sundar_2018, Tan_2017, Vu_2018, Xiao_2017, Zeng_2016} & \cite{Chen_M_2017, Fan_2017, Meng_2019, Ouyang_2019, Sundar_2018, Vu_2018, Xiao_2017, Zeng_2016} \\ \hline
	\end{tabular}}
\end{table*}

%% file: conclusion.tex
\section{Summary and Future Research Directions}
\label{sec:conclusion}

% Summary and key points of Table 2
We consolidated the literature based on our proposed taxonomy in Table~\ref{tab:classification_table}. 
As seen, with respect to the timing model, there are sufficient studies for both response time minimization and deadline constrained problems. However, there are limited works that minimize makespan. This is reasonable as makespan minimization is, in general, a harder problem to solve as the complexity is higher due to the inherent min-max optimization. In terms of contention, most contributions are on only computation contention and relatively fewer contributions consider both computation and communication contention.
Note that, the literature on no contention forms an interesting body of work since they mainly consider multi-objective optimization such as energy-delay trade-offs. From the perspective of problem classes, we find that there are very few studies that investigate all three problem classes combined: offloading, provisioning and scheduling. The existing literature primarily focuses on centralized solutions and there is little focus on decentralization. Further, among the decentralized solutions, very few works considered the deadline constrained timing model.

% Summary and key points of Table 3
An overview of how time-related model parameters (see Section~\ref{sec:architecture}) are used in the literature is presented in Table~\ref{tab:classification_table2}. As seen, there are fewer contributions towards multi-tier architectures. Only a few papers model queues on servers by considering serving rate and arrival rate of tasks. Lack of queuing models make it harder to address the multi-tier architecture problems. Most existing works assume the computation time of tasks on servers is known apriori, which may not be realistic. Finally, it can also be seen that compared to computation resource modeling, communication resources are relatively less explored in the literature.
Observe that only those papers that model bandwidth capacity have communication contention and those that bound the computation resources either in the form of computation capacity or serving rate have computation contention.

%Summary and key points of Combination of Tables 2&3
Comparing across Tables~\ref{tab:classification_table} and~\ref{tab:classification_table2}, we see works in both computation and communication contention category are majorly on single-tier architectures. Many multi-tier architecture works ignore the delay between servers. All contributions based on queuing theory consider only computation contention and provide only heuristic solutions. Interestingly, no surveyed work modeled queues and provided a decentralized solution.

% Future work/Outlook
From the literature, we observed that certain assumptions on problem classes and solution types leave some open problems. 
As discussed before, decentralized solutions with deadline constrained model are generally lacking. With the growth of decentralization in IoT applications, this is one potential research problem that needs to be addressed in the near future. 
Another important aspect to note is that most studies assume zero latency for the downlink data transfer (transmission of results from the cloud/edge servers to the devices). However, this assumption is unrealistic as certain AI applications (such as image/video search) have large data to be sent back to the devices. Although 5G technology offers higher downloading bandwidth, multiple tasks could contend for this bandwidth increasing the task response times.